# Security and Communication Network

# The Meeting of Acquaintances: A Cost-efficient Authentication Scheme for Light-weight Objects with Transient Trust Level and Plurality Approach

Tran Khanh Dang[1], and Khanh T.K. Tran[1,2].

[1] HCMC University of Technology, VNUHCM, Ho Chi Minh City, Vietnam.

[2] Can Tho University of Engineering and Technology, Can Tho City, Vietnam.

Correspondence should be addressed to Tran Khanh Dang; khanh@hcmut.edu.vn

## Abstract

Wireless sensor networks consist of a large number of distributed sensor nodes so that potential risks are becoming more and more unpredictable. The new entrants pose the potential risks when they move into the secure zone. To build a door wall that provides safe and secured for the system, many recent research works applied the initial authentication process. However, the majority of the previous articles only focused on the Central Authority (CA) since this leads to an increase in the computation cost and energy consumption for the specific cases on the Internet of Things (IoT). Hence, in this article, we will lessen the importance of these third parties through proposing an enhanced authentication mechanism that includes key management and evaluation based on the past interactions to assist the objects joining a secured area without any nearby CA. We refer to a mobility dataset from CRAWDAD collected at the University Politehnica of Bucharest and rebuild into a new random dataset larger than the old one. The new one is an input for a simulated authenticating algorithm to observe the communication cost and resource usage of devices. Our proposal helps the authenticating flexible, being strict with unknown devices into the secured zone. The threshold of maximum friends can modify based on the optimization of the symmetric-key algorithm to diminish communication costs (our experimental results compare to previous schemes less than 2000 bits) and raise flexibility in resource-constrained environments.

## Introduction

Advances in the wired, wireless, cellular and sensor networks have built a substantial infrastructure for the IoT. Sensors, RFID tags, smart thermostats, PDAs, smartphones, gadgets will be enabled to sense, process and control real-world events. Among the feature of IoT, security issues are an open and incredibly complicated. The three most important security requirements are authentication, confidentiality, and access control. Of these key security requirements, authentication is the first step when the system begins to operate its workflow. Nowadays, researchers have applied the authenticating into many different environments, the most concern are the energy consumption and the processing time of the participating devices which are limited resources: small-space storage, low battery's power and relatively short transmission paths.

With the technological advances of the sensor technologies rapidly, Wireless Sensor Network (WSN) has become the main technology for IoT. WSN is a mobile ad-hoc network in which sensors have limited resources and communication capabilities. They contain a large number of spatially distributed devices that are the low cost, ease of deployment, and versatility. Especially, limited battery energy and storage space are the tightest resource constraints on WSNs. These constraints caused the authenticating for mobility nodes more difficult. To be able to easily experiment with constrained objects, WSN is a main model of our scenario, which represents the IoT environment.

Earlier research works [1, 2, 3] have proposed an authentication protocol based on a Credential Service Provider (CSP) or a Trusted Third Party (TTP) to grant, distribute, and manage the key in an area or the whole system. These protocols consume a high overhead of exponential computations in resource-constrained environments. In the past the issue of mobility and how mobility can be used to boost authentication has been already proposed in the context of key establishment of wireless sensors network [4, 5]. Yet such protocol had a definite goal in mind: "*eventually all*" nodes should establish a security association with each other. Friendship or neighborship is known "*somebody you know*" which is a factor to help confirm a person [6] and reuse the closest objects discovery [7]. In this paper, we proposed a tightening authentication scheme incorporates a lightweight key exchange model through a behavior-based rating of mutual friends. Specifically, the main contribution of this paper is twofold:
1. Firstly, a novel authentication scheme which helps an unfamiliar object to join in a specific zone and to conduct communications over unsecured channels between objects is proposed. Our research focuses on the mutual authentication of individual nodes with the ability to move through areas within WSN. Since then, we combine the reuse of former interactions with the corresponding reputation points to derive a formula for calculating the average score and then compare it to a threshold value that is calculated depending on the process of setting up and the number of devices in the network. With this final result from the derived formula, the object which receives the initial request can give a decision to agree or refuse to make a connection with the coming object that wants to join in the WSN. Experimental results show the proposed scheme reduces the communication cost significantly on the transmission channel. This has reduced the delay time in the authentication process between the related devices compared with two recent mechanisms.
2. Secondly, a suitable key exchange scheme in conjunction with the above authentication process is introduced. Using the strengths of symmetric and asymmetric encryption helps key exchanging more securely and reduces costs instead of applying only one type of encryption in communication. The security analysis with BAN-logic indicates that our proposed security mechanism can resist typical types of attacks such as the impersonation attack, the message modification attack and the replay attack.

The rest of the paper is organized as follows. Firstly, we give an overview of the problems of research in authentication, trust-based and key management at section §*Related Works*. Then, section §*Approach* discusses the application scenario, assumptions and offers the security goals. In section §*Details*, we describe the detail of the lightweight authentication scheme. Section §*Evaluation* is divided into two pieces: subsection §*Simulation* shows the description of our simulation and CRAWDAD's dataset we referenced to rebuild a new dataset, and subsection §*Results* shows a detailed analysis of the found results. Security discussion will be presented in section §*Security Analysis,* it exposes security related issues: subsection §*Security Model* uses BAN-logic to provide the proofs which prove the correctness of the key exchange model in the authentication mechanism, threat model and security analyses will be presented in §*Threat Model* and then a brief metric for evaluation of two previous articles and our scheme will be detailed in subsection §*Security Feature Comparison*. Finally, we conclude the paper with a summary of our results and discuss opportunities for further research in Section §*Conclusion*.

**Related Works**

We also briefly compare the main features of the previous works to ours in the Table 1. In addition, we present our detail discussions about the advantages and unresolved drawbacks in this section. In terms of evaluation and comparison criteria, the communication cost that they need to ensure the cost of exchanging between two objects is not overwhelming while the authentication occurs, this is an important criterion because it affects the latency when an object waits for response and the energy-use in transmitting. These protocols also require an authenticating between the two parties and suitable to compare with our proposed protocol. Once we have the authentication, not only the related costs are taken into account, but also the secrecy of messages should be considered over an insecure channel. Besides, a key exchange mechanism between the involved objects must be investigated as well to establish the practical value of our newly proposed approach, e.g. how is the key exchanged or stored? do those keys depend on the central authority?, etc.

Localized combinatorial keying (LOCK) [8] is an Exclusion-based System (EBS) dynamic key management scheme and localized combinatorial keying. EBS is an exclusion-based system, a combinatorial formulation of the group key management problem. The system assigns each node k keys from a key pool of size *k+m*, where *k* is the number of administrative keys and *m* is the number of rekeying messages. In LOCK, nodes create a set of backup keys and only share with a base station. If a node is captured, the system will rewrite the key locally for other nodes. Moreover, if a cluster node is an intruder, a base station will be the main subject to rekey with other cluster leaders. LOCK is secured, can be backed up and shared the key to check with a Base Station (BS) and clusters when the network is attacking. When a cluster leader has been captured, LOCK deploy a new cluster leader to authenticate using backup keys shared with the base station. The communication cost depends on rekeying overhead. However, LOCK spent excessive amounts of time to wait for the rekeying from the BS to the other clusters. The related nodes are immobile so LOCK is not suitable for wireless environment like WSN. In addition, Chan and Perrig [9] built a key establishment scheme for one or more nodes as intermediaries to set a secure channel between neighbors. Using a node as trusted intercessors establishes a shared key between nodes in which each node has unique pairwise keys with $2\sqrt{n-1}$ nodes where *n* represents the number of nodes in the network. In KDC approaches, Peer Intermediaries for Key Establishment (PIKE) increased security performance and solved high density in the random key distribution mechanism (RKP) on logical grids. PIKE used the help of a trusted node to establish the pairwise key and exchange data on these keys. Computing and distributing keys are very crucial features in the protocol; yet they assumed that the sensor node is immobile; this approach does not good for dynamic context that we desire to resolve. We have considered some standard key management approaches to apply to our authentication mechanism. In addition, our scheme explored the PIKE's idea of using the help of a trusted intermediate node. With the intermediate node idea, we learned some evaluation of trust evidence was solved as a path on a direct graph, whereas nodes as objects and edges as relations between two objects. There was no pre-established infrastructure. Their idea was [10] also based on the direct trust relations of intermediate nodes. It took the weight of the edges to calculate trust values and establish an interaction between nodes. However, at each step of computation, the source node computes many times so the implementation may be complicated. They proposed a reputation-based framework [11] for sensor networks where nodes maintain a reputation for other nodes to evaluate their trust through metrics, represent past behavior of other nodes and be used as an inherent aspect for predicting their future behavior. We used these above reputation ideas for the algorithm of representing, updating, and evolution of trust in our scheme.

Table 1: Contribution of the previous works and our protocol.

|  | [9, 8] | [4, 5] | [3, 2] | [12] | [13] | [14] | Our proposed approach |
|---|---|---|---|---|---|---|---|
| Communication Cost | ✓ |  | ✓ |  |  | ✓ | ✓ |
| Authentication | ✓ | ✓ | ✓ |  | ✓ | ✓ | ✓ |
| Confidential | ✓ | ✓ |  |  | ✓ | ✓ | ✓ |
| Availability |  |  |  | ✓ | ✓ |  | ✓ |
| Mobility |  | ✓ |  |  | ✓ |  | ✓ |
| Key exchange management | ✓ |  | ✓ |  |  |  | ✓ |
| Establish trust relationship |  | ✓ |  | ✓ | ✓ |  | ✓ |
| No central authority |  | ✓ |  | ✓ |  |  | ✓ |

[13] created a temporary trust relationship to improve the authentication process which is also relatively close to our idea. Because they were the vehicular ad-hoc networks, the requirements were also appropriate to low the cost of computing in wireless networks. Once a valid vehicle is successfully validated, the protocol will become a valid vehicle for another vehicle, which is a temporary police car. The protocol ensured that it was small, anonymous, location privacy security; data could not be changed (upon the hash function). They presented detailed simulation parameters about the speed of vehicles and the size of network area. Indeed, they proved that their ability may apply this technique for the real scenario. About the CA, they did not mention about the trusted or base station to help their authentication and the process of exchange information between vehicles. Nevertheless, they did not care about the security of information exchange. [12] is also an authentication protocol with two-phase, because the first phase is registration phase so we are concern about the last phase, called authentication phase. In this phase, they used mutual authentication between client (U) and sensor node through Message Authentication Code (MAC). Besides, they applied symmetric key for the encrypted MAC on the unsecured channel and asymmetric key to verify MAC, then two sides create a trust relationship and get data from sensor node. They reduce the computational cost and energy cost in the authentication phase, using the hash function, symmetric and asymmetric keys. In addition, the protocol affects the optimization of encryption techniques, such as TinyECC. However, if it used for a moving sensor, it will be difficult because they depend on CA's public key from the registration phase. Because the key is managed by the CA, there is no key exchange management.

Broadcast authentication is a fundamental security service in distributed sensor networks, which rely on the hierarchical model including BS and member nodes. It uses key-chain as a secure storage to store mutual keys. BS generates the keys in each interval time (T) or in some specific cases then they are distributed to a group or all nodes in the network. Nonetheless, broadcasting consumes more energy over a period and has to deal with the security risk communication between nodes. Therefore, it is hard to determine T exactly for balancing between security and energy consumption. [14] proposed an authentication protocol for the pairs of nodes or between a node and a BS. The main point is they used a hash-chain to generate the key and the pair-wise key to swap. When these nodes communicate with each other, they utilize the hash of the key several times and add some parameters. The protocol is secure, yet it could not be used for the kinds of resource-constrained networks, because they proposed many mathematical operations and had to figure out the keys for each communication. In general, broadcast authentication is often used in distributed models, but the main disadvantage is that when participants, such as sensors or tiny devices, are in a harsh situation. These schemes cause slower response times; the cost of communication consumes so much energy. Thus, in order to solve this problem, our article focuses on peer-to-peer relationships, reducing the importance of central stations rather than go through the third party like a BS or a CA.

Capkun's research works [4, 5] are the closest to our idea. They proposed three mechanisms that support the establishment of new security associations. An interaction can be created by a set of materials, which includes user-name, its public key and its address on the secure channel. Later, they can verify that the node addresses match the public key. Although these mechanisms are simple and useful for the full self-organized network, they do not evaluate the responses of friends since they claim that those recommendations are correct. There is no evidence that the node is not fake, they did not consider the time of their neighborships. In addition, they do not have the detailed analysis of communication costs as well as key management process when the authenticating is needed. We had a short description of the authentication scheme through the voting and the reputation along with further discussion more about Capkun's model through a simulation algorithm in a brief article [15]. The authentication rate is more than 50%, the protocol is not too strict.

Amin and et al. [2] built a novel architecture suitable for WSNs. They designed a user authentication protocol and session key agreement for accessing data of the sensor nodes through multi-gateway. They managed smart cards and passwords to register and get access to a system. We do not deny the strength of this work but only offers a solution in the event that their gateway nodes are no longer usable. After receiving the login message from user $U$ who want to access data from sensor node $SN_j$, the HGWN (Home Gateway node) checks whether $ID_{SNj}$ exist or not. If the condition is correct, the HGWN will execute the authentication process; otherwise, the HGWN will send a broadcast message to all gateway nodes. It is confirmed that $ID_{SNj}$ matches with any one of the gateway nodes as foreign gateway node (FGWN) which executes the authentication phase. Their solution is secure and reduces the cost of energy, storage and time, but the components involved in the authentication process are the real user who comes with the smart device and the gateways, which get real-time data from the sensor nodes. Similarly, [3] created a novel remote user authentication scheme by using wireless sensor networks (for agriculture monitoring). The proposed scheme involves four types of participants: User/agriculture Professional $U_i$, a Base Station *BS*, a Gateway Node $GWN_j$, and Sensor Node $SN_j$ which user prefers to get some environment data. Instead of broadcasting to multiple gateways in multiple areas as [2], they would perform authentication via BS and GWN. Thus, we prefer to provide an additional solution that alleviates the communication cost when the scheme focuses on the role of *CA*, *BS* or *KDC* and guarantees flexibility, availability and tightening all the traffic moving in and out of the secure zone.

# Problem Solution & Proposed Approach

## Problem Statement

The explicit context is a motivation to improve and analyse the protocol properly in a wireless network. The characteristic we exploit is an ability to detect objects nearby and the objects are active in a specific zone. Currently, the IoT wireless network is widely available, and we are going to refer a very basic example that will be covered throughout the article. We will analyse the movement of humans/objects and the authentication when they meet in a zone/area or a meeting room. They will use a smart object to represent themselves such as a smartphone, a smart pen, etc. Because the meeting is secure, we will scrutinize they do enjoy the room and hidden their personal privacy. Besides, it is crucial to identify whether this newcomer is trustworthy or not. We will not use any intermediate helper center (*CA, KDC, etc.*) since it will spend a lot of time and be an inconvenience for small objects, portable and low battery.

Our objects can communicate with each other through wireless technologies such as *Bluetooth, ZigBee, wifi-direct, etc*. However, we do not assume that they are necessarily connected to the internet and can use the internet as an indirect connection mechanism. The signal transmission range of these technologies is very short as well as the bandwidth is also lower than the technology used for high power devices such as *Global System for Mobile Communications* (*GSM), Digital Telecommunications Cordless Telecommunications (Dect)*,...The number of members is large enough in a zone and must satisfy that when an intermediate usher explores someone. Because we apply a voting scheme in majority rule, our protocol can always make a yes or no decision. We do not assume that a room is on a fixed amount of land with unlimited range. In addition, we also consider the reputation of objects, which is a public attribute to raise the value of the voting process and make a decision.

In the previous works that we have studied, they proposed the neighbor discovery protocols and provided some related scenarios. [16] WILDSCOPE is a geo-referenced proximity detection system that provides direct contact information and acquires location information when and where a contact occurs, the location will be acquired when contacts are detected; configuration via a lifetime model and validation against the ground truth, proximity detection is depended on low-power wireless, known to be high. Neighbor discovery protocol guarantees that at least one of the nodes is able to detect the other in range, the maximum latency between the time when the nodes become in range ($t_0$) and the other time when one of the two successfully receives the other's beacon ($t_b$) is $t_b\text{-}t_0 \leq E$. To make the scheme that is symmetry, they used the detector sends a unicast message to the detected node upon receiving its beacon. Therefore, we assume that the objects have been installed the detection system - WILDSCOPE. When an object moves to another zone, WILDSCOPE will find the beacon they desire. Moreover, we also suppose that that beacon accepts and is able to help the newcomer.

In terms of security, we make some assumptions that consider proximity and authentication of mobile objects; new challenging assumptions are not familiar to traditional approaches to authenticate in the network. We start the list below with the assumptions:

- We do not assume that all honest nodes of the network are the friend forever, so if their time of the last encounter was too long ago or someone notices that he/she is a black object or their resource is not enough to store.
- After an initial set-up phase, nodes will move and might meet other nodes while gradually losing contacts with past nodes. There is a *variable geometry* of the connections so that at any given time a node might be nearby only to some set of other nodes. It might not have direct access to past friends or might have decided to discard them as friends after the above one mentioned timeout. (As a prototypical example, each guided tour requires a payment. So you may not want that a tourist hanging nearby might be re-authenticated after the tour is over). Suppose that when some objects will move out of the initial zone, the number of moving objects changes slowly (no strong fluctuation).

- Our protocol will operate based on the additional information after the network system is fully constituted at the beginning of time (bootstrap phase). Therefore, if any new object wants to get involved in the network, it will establish some materials to qualify to join the zone including zone access token, initialize list of friends based on the zone they are assigned. The adversary will attack the objects that have some interactions or leave out the safe zone.
- To disturb the traffic flow, we assume that the system can dynamically change the limit of the number of requests sent to a node. Depending on the speed of the wireless channel, the number of requests will change that can distinguish this is the spam or not.
- The guide, which is detected by the newcomer, is a good object.

**Proposed Solutions**

Authentication is a major issue to exchange information or access the zone services. Not everyone can enter or access certain documents in the authorized zone. A newcomer uses its former secure interactions that have the corresponding material to verify. If there is no one to help, newcomers will not be able to join. The authentication mechanism for a device is concerned with not only the safety and correctness, but also the promptness, availability and energy saving. We examined some previous authentication mechanisms for wireless devices in the sensor environment and the IoT environment generally. They focused their trust on the central distribution that is a reason why they both made the cost of authentication increase and consumed more energy. With the above motivations, we use human relationships to solve problems by evaluating rely on the reputation scores. With a real scenario which is in the authentication of secure meetings, our idea depends on the past interactions of the object, the reputation score shows the level of trust between two objects who have known each other, in the article does not mention the management of reputation points, only use reputation scores to authenticate. Like the voting process, the Guide is responsible for making the final decision from the feedbacks. Based on the results of our implementation, our protocol narrows the number of members entering the zone quickly and saves energy for moving subjects according to the assumptions we have mentioned.

The IoT systems are required to provide accessing services anytime. As long as it suits the context of our research, we can operate with new requests to join a specific zone. Initially, the format of the message must be correct. Then, some adjacent neighbours have been discovered enough to organize a short voting process (the parameters are limited to show the suitability and availability of the protocol, the results are presented in the experiment section). This article offers all sorts of hypothesis about security in which the most important is man-in-the-middle, impersonation attacks. The attackers will use any technique to falsify the original target, which will mislead the recipient. We develop a key exchange scheme to supplement the voting process with reputation score. A device sends a message packet with an own ID and an encrypted message which contains a shared-key associated with a specific sequence number. An attacker will not know two objects whom he is aiming in which stage they are or how many times they communicated. It will initially be locked with a private key, and then two nodes will exchange a special shared-key depending on their order communication. Our key exchange model supports encrypt all messages and identifies the device's friends through their shared materials. This model is applied to each node without any trusted parties. In the authentication process, each device has just a unique ID to represent itself and is possible to detect nearby devices.

## Protocol Details

We build the idea of the organization and the operation of the meeting room, staffs use a smart device to identify and join a secure zone. However, we use the trust relationship of nodes (all the neighbors at the same level), a reputation-based method for each small device. These devices do not rely entirely on the trusted third parties to alleviate the cost of communication, the processing time and then ready to allow nodes' authentication anytime.

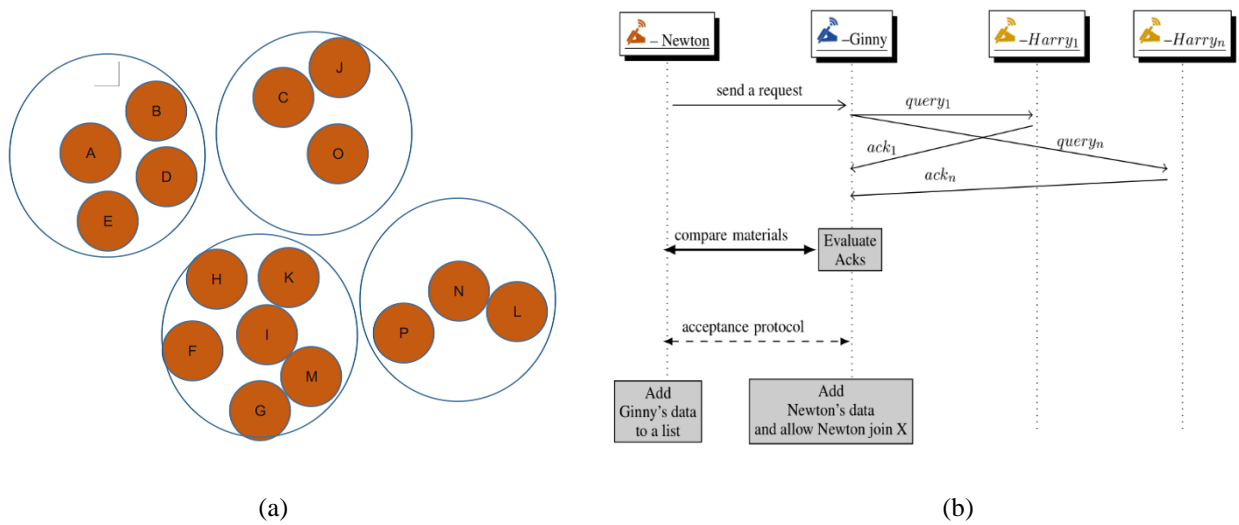

(a)  (b)

Figure 1: Overview and the Enhanced Authentication Process for Object's Encounters in the Meetings

**The Familiar Voting Authentication Scheme**

Figure 1a describes the overview of our protocol. In Figure 1b, we introduce three main roles: a newcomer (Newton), an intermediate guide (Ginny) and some helpers (Harry). Firstly, Newton is an unknown person, moves close to the zone's bound and detects one of the adjacent objects by a discovery mechanism. The selected object as the intermediate guide will reach a request to make a secure association. Then, it will perform some operations to make a final decision, which allows Newton to take part in this zone by his previous relationships and his permissions.

From our assumptions in section Scenario, this protocol will be utilized after the system has organized completely (all honest nodes meet and have a secure interaction and distribute the zone's permission). At this time, each node contains:

- $ID_O$: a unique *ID* of each object *O*.
- *(S,P)*: a set of private keys and public keys.
- *ls_fr(O)*: a list of nodes which *O* has an interaction. It has a key-list (the list of sequence-shared keys that *O* received a message for the last time) and reputation scores, respectively.

- *ls_zone(O)*: (the zones that *O* is allowed to move to – assigned early): a list of zone's *ID*.

We describe the steps of the mechanism below and use the specific name of the person to represent the respective components.

*Moving and Joining:* After the first step, the objects will get enough information to join the given zone. Newton moves to zone X and sends a request to enter zone X to Ginny who is detected the nearest one by WILDSCOPE. At this point, Newton satisfies the conditions: a zone's permission and must be Ginny's friend or be a friend of the nodes who is currently in zone X. *[has_permission(X) and is_friend(Ginny)]* or *[has_permission(X) and is_friend(some of all zone's member)]*. We distinguish two cases in this step:
- *Are you guide's friend?* From the *is_friend(Ginny)* function, we consider that Newton's ID exists in her friends' list or not? From the very beginning, Newton had discovered Ginny and sent a request to the zone. Because Newton and Ginny established a relationship, he would also enclose the shared key as a secret key between the two objects with the message to interact. From there, Ginny can easily check the message content and evaluates whether to enter or not.
- *Are you neighbours' friend?* If Newton does not know Ginny, Newton will only send his permissions and ID to her. After that, Ginny explores the closest objects around her and sends the messages, which expect some helps to verify. Our protocol regulates not only the number of nodes in/out to the zone but also the choice of helpers. It will happen in two ways: firstly, Ginny finds the appropriate number of helpers and a verification message should be sent to the corresponding helper sends, and then, if there are too little or no more helpers around, Newton will not be allowed to enter into zone X.

*Checking and Response*: When the helpers receive a request from Ginny, they execute the is_friend(Newton) to detect a previous interaction with Newton. In order to diminish the flow of messages during the authentication process, helpers will not need to send back Ginny if they do not have a preceding association with her? Here, Harry will be described as a helper to illustrate the task, [Newton is my friend]: Harry will send including Newton's ID, Harry's ID, and Newton-Harry's shared key.

*Evaluation the Acknowledgements (ACKs):* The received *ACKs* from the Harry(ise), Ginny will selectively filter the types of message by the reputation score. Because Harry(ise) is(are) all currently Ginny's friends and has(have) dealings with her in zone *X*, she will have the respective reputation values, thenceforward, we may have the average Newton's reputation from Harry(ise).

$$Total\_trust(H, C) = \sum_{n=1}^{h} \frac{M(G, H_n) * R(H_n, C)}{m} \qquad (1)$$

where:
    *H*: a list of helpers, *H = {Harry$_1$ ,Harry$_2$ ,...,Harry$_n$}*
    *C*: a newcomer object.
    *G*: an intermediate guide.
    *h*: the number of neighborhoods who *G* detected.
    *M(G,H$_i$ )*: a value of *H$_i$*'s multiplier by *G*.

*R($H_i$,C):* a value of *C*'s reputation score that $H_i$ stores in its space.
*m*: the sum of *M(H,C)*.

Before evaluating the value of *T(G,C)*, Ginny will ask Newton to compare shared materials which will prove actually this association was created. Then, we set up a scoring system and a multiplier for the reputation score. *Total_trust(H,C)* (1) is the average of aggregate value which Ginny calculated from the ACKs of Harry(ise) on the newcomer (Newton). The scoring system creates points based on the behavior and the number of members over time; we need to choose a limited score or a maximum value that is large enough for the reputation of each node. To make it easy to identify trustworthiness, we classify three different levels of reputation: 0: too bad [0], 1: Fine [1,$\frac{N}{2}-1$], 2: Good [$\frac{N}{2}$, *N*]. *N* is the maximum value of reputation that the system needs to establish before the authentication scheme performs. In the experiment section of this article, we took *N* as *100*. These levels are the multiplier we will use to raise or lower the value of trust from the Harry. At this point, Ginny will evaluate the reputation of Harry(ies) in one of the predefined value ranges of the above levels to deduce the corresponding multiplier. For example, if Harry's reputation score that Ginny rated as *4* and *N = 10*, Harry's multiplier by Ginny will be *1*.

$$NA(H,C) = \frac{Total\_ACKs}{Total\_ToDeliver} \qquad (2)$$

$$T(G,C) = \frac{NA(H,C) + Total\_trust(H,C)/N}{2} \qquad (3)$$

where:
*NA (H, C)*: The ratio of the positive number of *H* about *C* over total number of queries has been sent.
*Total_ACKs*: the total number of acknowledgements received.
*Total_ToDeliver*: is the total number of messages delivered.
(0 <= *Total_ACKs* <= *Total_ToDeliver*)
*T(G,C)*: an average value of trust about newcomer by guide.
*N:* a maximum value of reputation.

This article investigates a novel authentication scheme rely on the secure voting mechanisms such as human-to-human relationships without the significance of the trusted third parties. We are interested in two types of value: *Total_ACKs* and *Total_ToDeliver* (2) will consider the number of friends *H* positive *ACKs* about *C*. Then, we proposed an algorithm based on (1) to build a combined evaluation result from Harry(ise). Evaluation of each node's individual reputation scores produces a subjective result, so we will consider the number of responses received as the votes for Newton to balance the final result. (3) is an association of a ratio of the positive responses and a ratio of the average reputation score, is a final result which Ginny needs to make a decision on Newton's join request. The value of reputation ranges from *0* to *N*, where *0* is computed in (1), it is as an empty vote when we evaluate the number of ACKs. Besides, we add a threshold of values for trust-related problems is Δ$_{reputation}$ which we compare to the value of *T(G,C)* in (3) which is a trust average value of newcomer by guide. If *T(G,C)* ≥ *Δ*$_{reputation}$, a new association will be established, both of Newton and Ginny will store their information then he can take part in zone *X*. While the system is operating, the algorithm computes the reputation points on each node that always be activated. In order to build algorithms flexibility, we will not spell any specific behavior out in this article. The value of reputation (two nodes are friends of each other) will raise or reduce against the communication status between two nodes. We divide it into four cases including sending-receiving (in a friendship), evaluation, suspicion and attacks. If sending-receiving or evaluating is successful, the reputation score will go up by *1*. Otherwise, the score will be reduced to *1* based on their behavior in the suspicion cases or attacks.

**Key Exchange Scheme**

Key exchange scheme is a mechanism by which two objects that communicate over an adversarial controlled network can produce a secret shared material. Key exchange protocols are essential for enabling the use of shared-key cryptography to protect transmitted data during an authentication process. In this article, we propose a key exchange mechanism to match the constraint requirements of participants. Our technique can not only save time but also ensure the forwarding data are secure. As we have studied the symmetric and asymmetric encryption, both of them can support for resource-constrained devices, but there are drawbacks. So that, we combine their advantages to increase security and slightly down processing time.

As illustrated by the example on Figure 2 below, we will use the same object's name. The scheme consists of two stages: first-meet and meet-again. In the first phase, Alice uses Bob's public key to encrypt a packet, which is included:
- *$ID_A$*: determines who is sending a message for Bob.
- *$Snum_B$*: a sequence number of Bob. At the first time, *$Snum_B$* = $_0$, after Alice has sent a message to Bob, *$Snum_B$* will increase the value by one when Bob did decrypt Alice's message successfully to determine the correct order of messages, also to avoid the attackers which send fake messages to Bob.
- *$K_{ABi}$* : a key to decrypt for the following message that Alice will send to Bob (with i is a version number). This key will be attached to the original message.
- *$E_M$* : an encrypted message which combines an message and *$K_{ABi}$* . *$E_M$* (M) is a function to encrypt a message M with a key named *K*.
- *$De_K(E_M)$*: a function to decrypt an encrypted message *$E_M$* with a key named *K*.

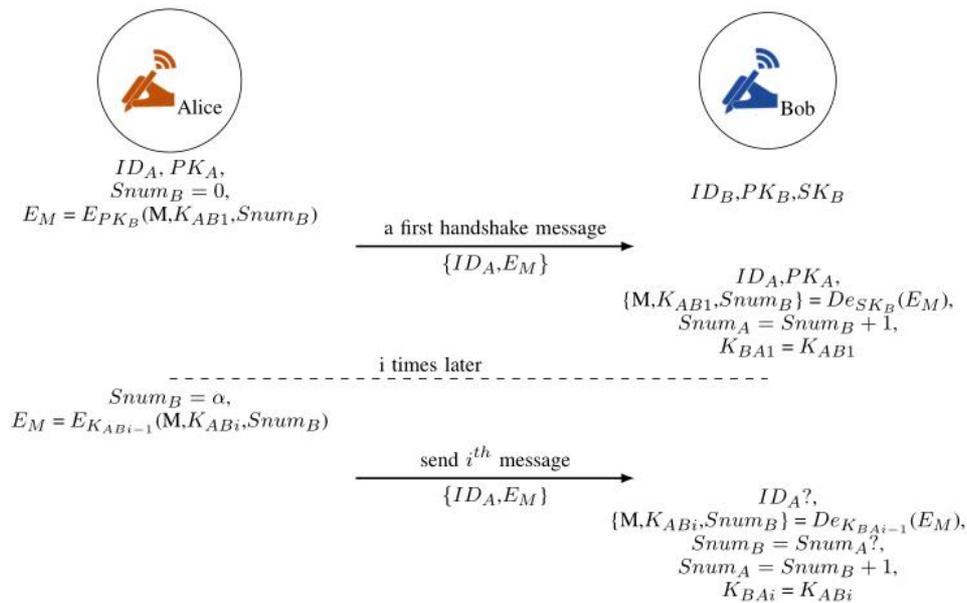

Figure 2: Key Exchange Scheme

Alice has an original message (M), creates a shared-key ($K_{ABi}$) and an initial sequence meeting number ($Snum_B$). Afterward, Bob has received an encrypted message ($E_M$) from Alice, he will apply his private key to decrypt $E_M$ with a function as $De_K(E_M)$. Subsequently, Bob will get an original message (M), $K_{ABi}$ that decrypts the next exchange message and $Snum_B$ which Alice sent Bob to check for the exchange later. At that time, Alice also stores $K_{ABi}$, $Snum_B$ in her private memory. We use the asymmetric cryptography once at the initial exchange to help make the scheme's security safer. We assume that when Bob and Alice meet again after a period, she sends a message to Bob at $i^{th}$ (with $i \geq 1$). Because Bob has had the shared key ($K_{ABi-1}$) from the previous communication with Alice to decrypt the current message, Alice will use the key she gave Bob at the $i$-$1^{th}$ to encrypt the $i^{th}$ message and build a new shared-key to send to him with the above same information. After Bob has received the message and decrypted it successfully, he will check $Snum_B$, which represents the number of times they have interacted with each other. If it is valid, he will store the key and $Snum_B$ that Alice has just sent, increases $Snum_B$ by *1*. At the same time, Alice and Bob will keep the new shared-key to replace the old one, a new *Snum* and $ID_A$ is not changed.

We are based on the idea of creating the key from the scheme which called μTESLA [17] that has been proposed a broadcast authentication in a distributed sensor network that is improved from a stream authentication protocol called TESLA [18]. μTESLA uses a sequence of authenticated keys that are continuously linked together by a random function that is a one-way function. Each key in the key chain is the image of the next key in random function. In our paper, the initial interaction uses asymmetric encryption; and in the subsequent interactions, we will use symmetric encryption; the key of these interactions is generated from the image of key in the last connection.

## Protocol Evaluation

**Simulation**

To validate our protocol, we referenced the mobility dataset from CRAWDAD collected at the University Politehnica of Bucharest in the spring of 2012 and applied a dataset entitled HYCCUPS Tracer [19] with the purpose of collecting contextual data from Android smartphones. They gathered information about a device's encounters with other nodes by using AllJoyn. This software will continue to be available for developers to download and run on popular platforms, especially for many another lightweight real-time operating systems without the need for internet access.

Table 2: Experiment environment

| Parameter | Value |
| --- | --- |
| OS | Window 10 Pro 64-bit |
| CPU | 2.4GHz |
| RAM | 4.0Gb |
| Programming Language | Java |
| Libraries | gson-2.2.2, apache.common.io-2.6, JRE1.8 |

We based on the format of an original script from Hyccups database to exclude and append the properties, objects and lines. We have 100 objects as sensor nodes. A list of friends and a current zone generated randomly to simulate the initial phase when the network was stable. Our experiment runs an algorithm for mobility devices in nearby areas. It creates a local movement within zones (from 1 to 19). The big result we reached 99999 connections. Since the period will greatly affect our results, the dataset will not concentrate on the time when they interact.

| No. | Value |
|---|---|
| 0 | "{"node":14,<br>"friend_list":[63, 85, 67, 95, 55, 78, 43, 41, 13, 56, 68, 49, 8, 22, 81, 61, 28, 16, 96, 26, 19, 20, 65, 79, 17, 21, 9, 52, 92, 93, 89, 82, 72, 70, 12, 97, 3, 25, 84, 47, 51, 40],<br>"zone":[8, 12, 10, 4, 6, 19, 15],<br>"current_zone":17}" |
| 1 | "{"node":8,<br>"friend_list":[8, 54, 63, 1, 32, 7, 46, 81, 60, 65, 49, 77, 97, 16, 91, 86, 80, 13, 18, 71, 83, 52, 23, 28, 98, 92],<br>"zone":[18, 9, 19, 14, 12, 2],<br>"current_zone":3}" |
| 2 | "{"node":80,<br>"friend_list":[33, 73, 74, 97, 77, 81, 63, 100, 64, 7, 10, 61, 53, 4, 36, 16, 30, 8, 94, 96, 32, 54, 65, 79, 62, 67, 89, 90, 99, 80, 15, 50, 6, 78, 91, 85, 56, 3, 55, 83, 44, 98, 70, 58, 11, 48, 24, 17, 42],<br>"zone":[14, 12, 9, 1, 10, 6, 18, 20, 4, 3, 17, 16, 19, 5, 13, 15, 11, 7],<br>"current_zone":8}" |
| ... | |
| 97 | "{"node":77,<br>"friend_list":[9, 11, 28, 46, 4, 13, 75, 88, 48, 89, 94, 40, 41, 37, 25, 74, 76, 6, 20, 96, 47, 82, 35, 52, 57, 84, 100, 99, 78, 50, 80, 62, 33, 67, 56, 2, 51, 98, 32, 86, 1, 68, 59, 24, 22, 55],<br>"zone":[15, 20, 1, 12, 4, 16, 13, 17, 8, 14, 2, 11, 9, 18, 6, 7, 5],<br>"current_zone":3}" |
| 98 | "{"node":56,<br>"friend_list":[87, 40, 25, 98, 60, 46, 15, 85, 71, 34, 1, 96],<br>"zone":[7, 6, 10, 12, 2],<br>"current_zone":4}" |
| 99 | "{"node":87,<br>"friend_list":[18, 99, 40, 53, 37, 52, 3, 1, 63, 33, 5, 47, 41, 15, 59, 65, 12, 48, 23, 57, 64, 75, 32, 61, 54, 19, 92, 24, 55, 82, 7, 91, 78, 46, 4],<br>"zone":[16, 4, 20, 7, 11, 1, 10, 13, 18, 5, 17, 6],<br>"current_zone":14}" |

(a)

| No. | Value |
|---|---|
| 0 | "{"newer":67, "checker":8, "current_zone":3}" |
| 1 | "{"newer":70, "checker":74, "current_zone":13}" |
| 2 | "{"newer":39, "checker":99, "current_zone":8}" |
| ... | |
| 99998 | "{"newer":56, "checker":96, "current_zone":4}" |
| 9999 | "{"newer":51, "checker":19, "current_zone":2}" |

(b)

| Object | Attribute of an Object | Value of an Attribute |
| --- | --- | --- |
| Node | Node | ID of object; ID ∈ [1..100] |
| | friend_list | a list of friends |
| | zone | a list of zones where the object is allowed |
| | current_zone | node is in this zone |
| Script | newer | ID of a newcomer |
| | checker | ID of a checker |
| | current_zone | newcomer need to join this zone |

(c)

Figure 3: Dataset description: (a) List of moving nodes, (b) Experiment script, and (c) Explanation of the attributes of each JSON object.

Assume that each line will appear at different times and gradually increase from top to bottom in our script. As we mentioned, our protocol has two phases: bootstrap, moving and joining.
In the first phase, we set up information for an object, friendships (taken from the objects' dataset), zones, and grant zone's permission to the objects. Our project has two sections: creating a dataset of nodes and script, and illustrating. In the first section, we create some information for 100 objects, including their own ID, friend_list, zone (a list of zones where the object is allowed) and a current zone where they are. Then the last one is all script that we need to run the whole process of illustration. Our protocol will work with two types of files: *script.json* contains the movement script of the whole system (see Figure 3b), *nodes.json* is a list of the objects' information, the file, which be simulated as a storage space of an object (see Figure 3a). Each line of *script* file represents a specific time of an encounter with the first element (the left ID) as Newton and the second one (the right ID) as Ginny. Both of IDs are the same line when Newton meets Ginny. Besides, we add a Ginny's current zone attribute each line and a list of nearby neighbors that Ginny has traced.

The algorithm will start running after the end of the bootstrap phase, each time Newton goes to zone X, detects the closest one to him and sends a request with a certificate to join the zone (if any) because we want to reduce the confusion, we make the zone's rule fixed. Ginny gets the signal, checks the zone certificate for real or not and their interaction. Each node not only has a specific grant but also requires prior interaction for authentication. Table 2 gives the basic environmental parameters for our experiment. Each change of the parameters, the average execution time of experiment algorithm is approximately 425000 milliseconds (ms) with 99999 connections. It means that when two objects met up, the process would last for about 4.25ms.

In the previous sections, the protocol confirmed the goal of reducing the energy and the execution cost of the authentication process. Thus, we decided to choose TelosB as the platform simulation for sensor objects in our experiment. MEMSICs TelosB Mote is an open-source platform based on a low-power MSP430 16-bit with 10KB RAM and 48KB program flash memory. TelosB has a low current consumption and is powered by two AA batteries. It runs TinyOS and an IEEE 802.15.4 radio with an integrated antenna, which claims outdoor the range of 75m to 100m and indoor range of 20m to 30m. Besides, we could set for TelosB with a technology that detects the proximity sensors within its range coverage. Recently, TelosB is a promising sensor platform for wireless sensor networks as well as large networks such as *IoT*. A study [20] presented an energy comparison in the two states of idle and processing for all architectures. Overall, the results showed that TelosB is less energy than other platforms. Although the processing power of the MICAz (0.025mW) is lower than the TelosB (0.035mW) in the WSN environment, the MICAz (26mW) sleeps is up to 5 times the TelosB (4.8mW).

Table 3: Energy Cost

| Operations | Energy cost ($\mu J$) |
|---|---|
| Transmit 1 byte | 5.76 |
| Receive 1 byte | 6.48 |
| AES-128 bit | 9 |

US NIST announced a call for candidate ciphers for its new Advanced Encryption Standard (AES). After exhaustive analysis and evaluation, Rijndael was selected as AES finalist. AES is relevant to our goal, which is cost savings and energy. The AES cipher is obtained by running a number of transformation rounds repeatedly that convert the plaintext (original) input into cipher text (encrypted) output. When the plaintext is required, similar steps are carried out in reverse fashion to attain the plaintext. AES operates key-alternating block ciphers. The XOR operation is performed on the input array for encryption. Therefore, we use the AES to represent the encryption algorithm in the key exchange scheme. It makes our mechanism to compare to previous protocols reasonably.

**Results**

Based on the below chart, it shows that our protocol is strict when we should accept any node that joins the zone. The parameters that e set to find the appropriate threshold values dynamically. They are $\Delta_{max\_friend}$ and $\Delta_{reputation}$. Firstly, the number of friends in the same zone is important as it affects the energy of the guide. $\Delta_{max\_friend}$ is the maximum number of targets that the guide is able to send a query. Then, a threshold of reputation ($\Delta_{reputation}$) is a comparable value to make the final decision. We set up the parameter, which is an arithmetic progression with common difference of *1* with the following conditions: *$3 \leq \Delta_{max\_friend} < 100$,* there are *100* active objects in *19* zones (based on the sample data we have refined, there are some differences from the original data of CRAWDAD resource). In addition, $\Delta_{reputation}$ (*$50 \leq \Delta_{reputation} < 100$*) is the threshold of reputation after the guide receives assistance from nearby neighbors (because when evaluating a node's request, we will must calculate based on majority rule and the value must be greater than 50% which is valid, just like an election).

As far as the best data we received, the result of successful authentication is approximately 22% of the total number of nodes, which want to enter a specific zone. In addition, asymmetric encryption is only used once in the first time of interaction. It indicates that the performance is improved and saved the energy by transmitting messages between nodes in the network. Our scheme observes the number of sending and receiving information (see above, Table 3) which makes power and time can be calculated and derived during authentication efficiently [21].

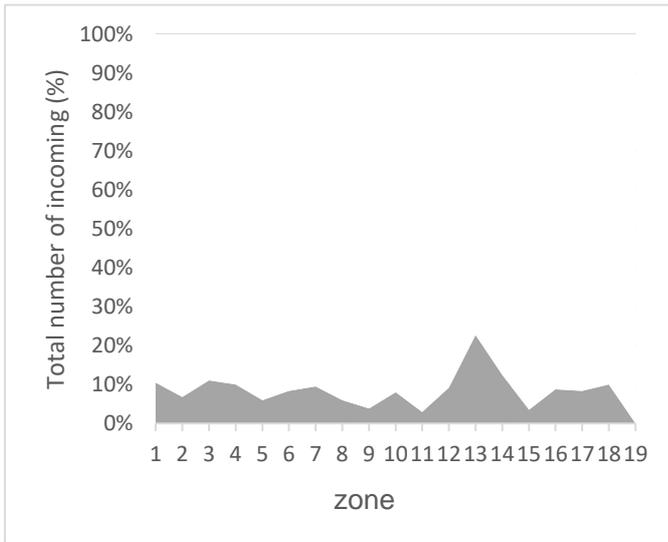

(a)

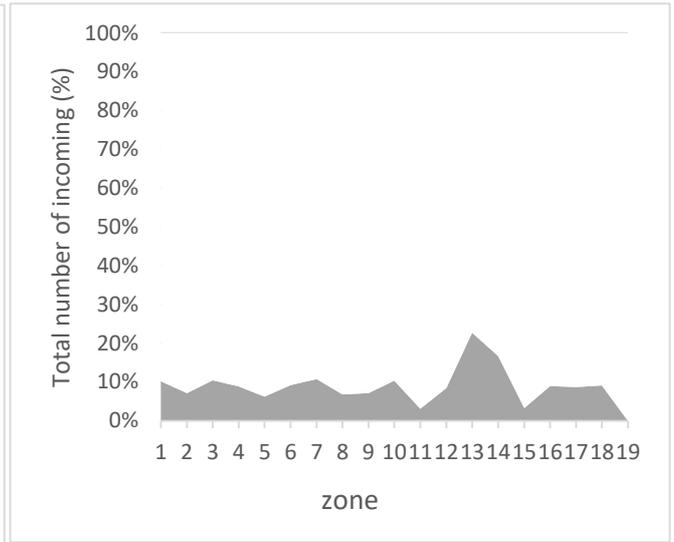

(b)

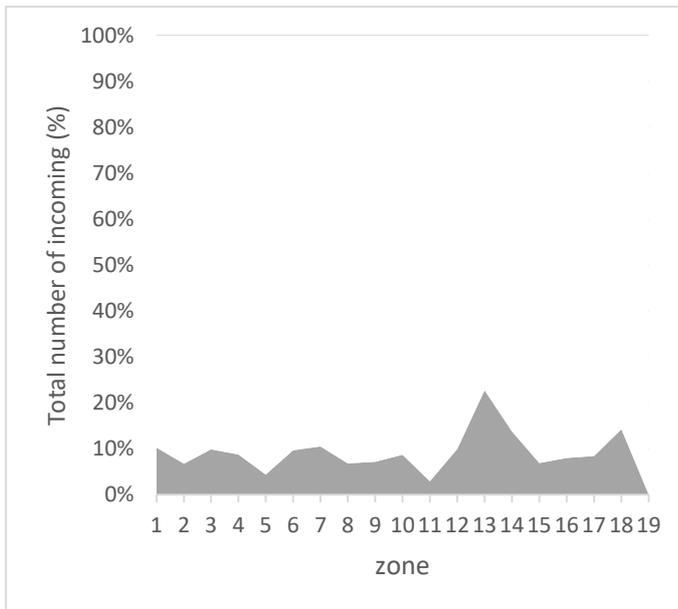

(c)

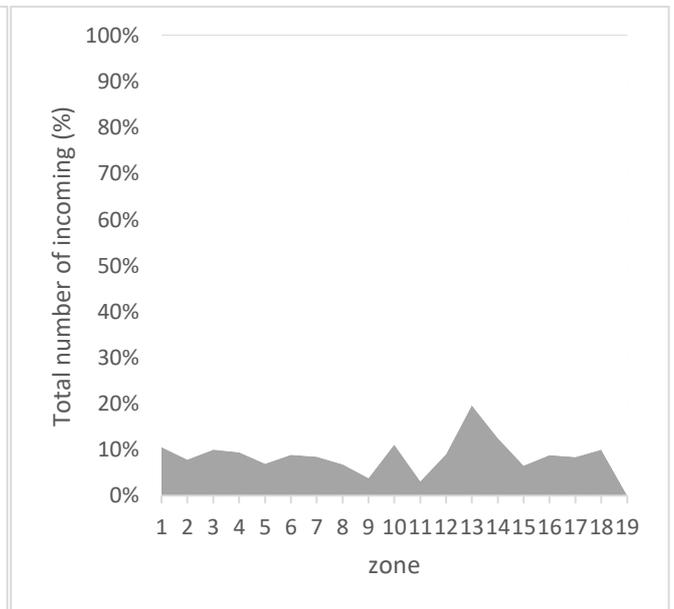

(d)

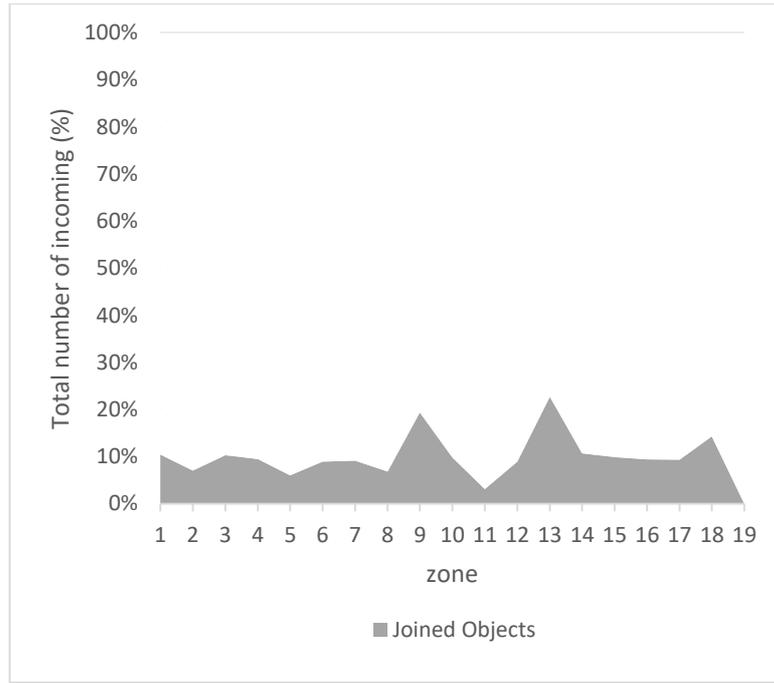

(e)

Figure 4: The number of incoming objects according some defined parameters with $\Delta_{reputation} = 80$
(a) $\Delta_{max\_friend} = 3$, (b) $\Delta_{max\_friend} = 7$, (c) $\Delta_{max\_friend} = 11$, (d) $\Delta_{max\_friend} = 28$, (e) $\Delta_{max\_friend} = 70$.

Based on the estimated values, we will evaluate the energy consumption of the proposed protocol and key exchange mechanism. In the Table 4, we show all cases occur together with the largest number (worst case) of all types of messages used. We experiment on the change of the total cost of our mechanism through the variations of $\Delta_{max\_friend}$ from *3* to *100*. The percentage of the number of accepted members into the active zones relies on the change of parameters is negligible from ~ *3% - 6%* (Figure 4). There are two situations: Figure 5 keeps a constant value of Δreputation is *80* in *[50,100]* and vary the value of $\Delta_{max\_friend}$; Figure 4 we have a constant value of $\Delta_{max\_friend}$ is *7* and vary the value of $\Delta_{reputation}$. Because we need to analyse the difference as well as the influence of these two properties. After that, we observe that the graphs are stable when the values of $\Delta_{max\_friend}$ are changing, and the graphs fluctuate when the values of $\Delta_{reputation}$ are changing. The two thresholds we define ($\Delta_{max\_friend}$ and $\Delta_{reputation}$) are two evidences that prove our proposed formulas and algorithm are stable and inspect the entry requests. With $\Delta_{max\_friend}$, it is a threshold of the maximum friends the node can ask for. There are five prominent variants of the all results in the simulation algorithm we list below (see Figure 4 a, b, c, d, e) when the values of $\Delta_{max\_friend}$ were change which is easy to infer that the authentication mechanism is very strict. Moreover, the results also show that the percentage of successful authentication does not change significantly and achieves the initial purpose while the number of friends increases or decreases. After that, when we combine the selection of the five above variants with the comparison of the previous proposed mechanisms, we have chosen the average threshold, $\Delta_{max\_friend} = 7$ for the latter experiment ($\Delta_{reputation}$ changes and keep the value of $\Delta_{max\_friend}$). $\Delta_{reputation}$ is the reputation score, when the final result that the guide (as Ginny) receives will compare to this value. If the result is greater than or equal to $\Delta_{reputation}$, the subject will be validated in the zone. The reason for this difference is because the value of the reputation is more influential than the number of voting friends in our formula, the behaviour and evaluation of an object will lead and affect the result properly. In addition, we also show two graphs (see Figure 5 a,b) that show the contrasts when changing the values of $\Delta_{reputation}$.

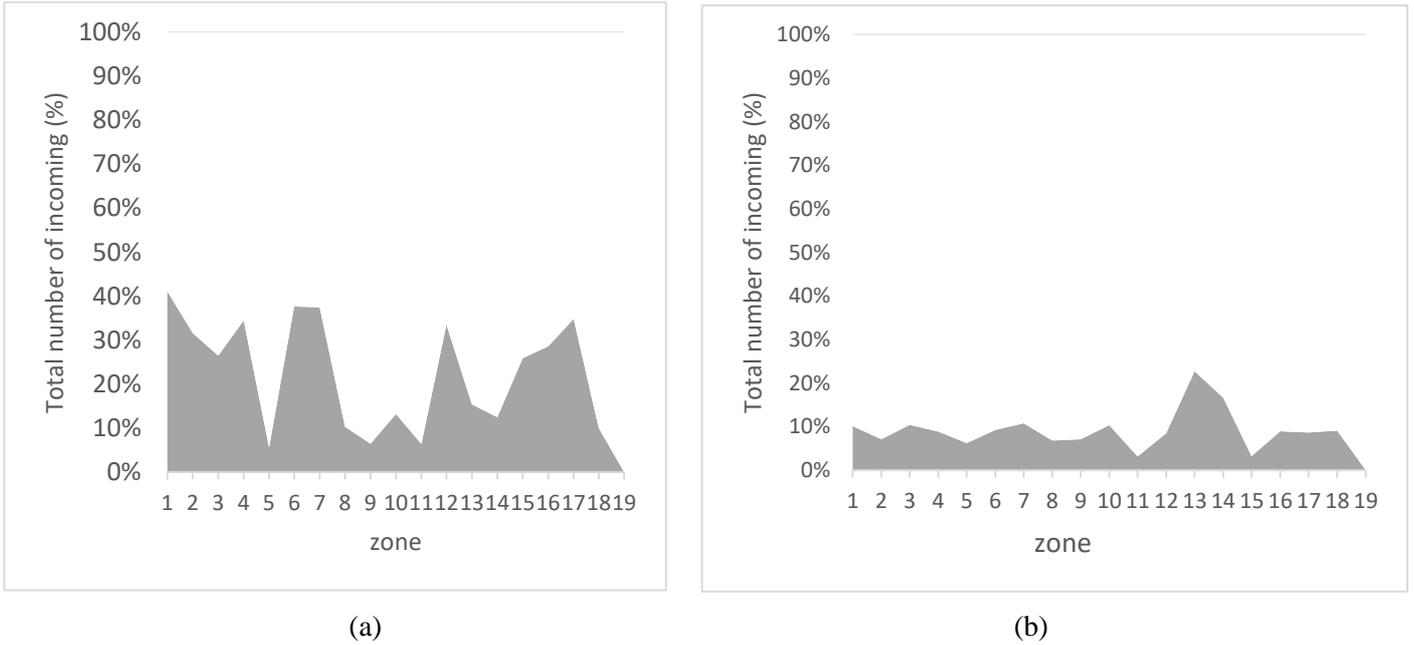

Figure 5: The number of incoming objects according some defined parameters with $\Delta_{max\_friend} = 7$
(a) $\Delta_{reputation} = [10,30]$, (b) $\Delta_{reputation} = 80$

The comparison result of the authentication process with the precious works from a mean value is *7*. Table 4 indicates scenarios, which get the *mean value* of $\Delta_{max\_friend}$. From here, we can calculate the maximum cost each a sensor object based on the above assumptions and authentication process. Case *#1* which Ginny and Newton are friends, Ginny will receive just one request and send a successful notification back to him. Case *#2* which Ginny asks her neighbors but they do not know about Newton. In addition, in case *#3*, although all Ginny's neighbors know Newton but the average reputation is not enough to let him join this zone. In case *#4* in this table, Ginny (as a guide) is a mediator that spends more time and energy than the others, so we focus on Ginny's message flow. Ginny is supported to run on the TelosB platform (the configuration mentioned above in this section); she will receive a request (16 bytes), a list about the shared information of Newton's friends (16 bytes). Besides, Ginny will send *7* queries to ask her neighbors who she could be found to support and can get back 7 ACKs. After that, if there is a validation result, Ginny will send a confirmation to Newton. From messages and the value from the Table 3, we can deduce the maximum energy consumption of each sensor as a guide that the protocol has gotten in all scenarios: $(7 * (16 + 2)) * C_{transmit} + ((7*16 + 2) * C_{receive} + 7 * C_{AES} * 4) = 1521\ \mu J$.

Table 4: The number of messages from our experiment

| Scenario | Transmit | Receive | Queries | ACKs |
|---|---|---|---|---|
| 1. They are friends. | 1 | 1 | 0 | 0 |
| 2. Do not have mutual friends. | 1 | 0 | 7 | 0 |
| 3. Have some mutual friends but fail. | 1 | 0 | 7 | 7 |
| 4. Have some mutual friends and success. | 2 | 2 | 7 | 7 |

In the Table 5 below, we calculate the computational cost for the whole process from the time of sending a request to join to the successful verification of a sensor node with the related participants. According to the experimental results, the encryption time for symmetric encryption (AES-128bit) from [3] is *0.130ms* and the execution time for hash function is *0.0004 ms ≈ 0.001ms* : Newcomer will send two hash messages includes a request and a list of mutual friends with an ID to the guide; we take the result in case *#4* in the Table 4 to compare so the guide sends *7* requests to helpers encrypted with the secret shared-key and the helpers unlock this encrypted message by a shared-key with a decrypt function, after that each helper send a reputation back to the guide (= *7*2+7*2 $T_S$ = 28$T_S$*). Finally, the guide has a final result for newcomer's request and notice it to newcomer. Therefore, we compute that the average execution time is *3.6 ms*. In this regard, our mechanism is slower than [3] about ≈*2ms-3ms*. We have to use multiple queries and send encrypted messages to neighbors rather than to rely on the base stations or central authorities or gateway nodes like [2, 3] which gives our scheme more flexibility. In addition, we also compared the communication cost for all objects in an authentication process to the above works. For calculating the communication cost, we consider the length of IDs, cipher messages, requests, confirm messages, and so on. Table 6 shows the comparison of the total communication cost with [2, 3], this result is calculated from case *#4* (see above, Table 4) combined with the participants in the entire process with *8* times of hash (160 bits/message) and *28* times of using symmetric encryption (128 bits/message) = *8*160 + 28*128*.

Table 5: Computational cost comparison.

| Schemes | Newcomer | The guide + Helper | Total | Execution Time (ms) |
|---|---|---|---|---|
| [2] | $8T_H$ | $7T_H$ | $20T_H$ | 0.008 |
| [3] | $6T_H + 2T_S$ | $(5T_H + 2T_S) + (8T_H + 3T_S) + (5T_H + T_S)$ | $24T_H + 8T_S$ | 1.052 |
| Ours | $2T_H$ | $2T_H + 28T_S$ | $4T_H + 28T_S$ | 3.64 |

$T_H$ is the execution time of the hash function; $T_S$ is the execution time of symmetric cryptographic algorithms.

Moreover, Table 6 also indicates that our scheme takes far less communication cost than the existing protocols. In general, though, our mechanism cannot optimize for both computational and communication issues, we lower the number of requests and components involved in the authentication process greatly while the computational cost we consume is only slightly slower. In addition to comparing computational and communication costs, we have a comparison of security features that includes known criteria and new criteria (passwordless authentication, no central authority and no key exposure) in sub-section of Security Analysis.

Table 6: Communication costs comparison.

| Schemes | Transmitting/Receiving | Total (bits) |
|---|---|---|
| [2] | 3776 / 2880 | 6656 |
| [3] | 3680 / 3840 | 7520 |
| Ours | 2432 / 2432 | 4864 |

## Security Analysis

**Security Model and Proof**

BAN-logic [22] is a beneficial tool for formally analysing an authentication scheme. Inspired by the analysing methods proposed by [23, 24] to evaluate our proposed scheme, we first briefly introduce the model including goals and assumptions, BAN-logic. We present our proof based on Burrows' BAN-logic model to evaluate the ability of an authentication scheme formally. According to BAN-logic, there are three objects in logic: principals, encryption keys, and statements (we will identify a message as a statement in logic). Therefore, the symbols P and Q range over principals, X and Y range over statements, and K represents the cryptographic key. Below are notations used in BAN-logic model.

- $P \mid \equiv X$: It means that P believes that in the current run of the protocol that the formula X is true.
- $P \triangleright X$: P who can read, receives a message including X.
- $P \mid \sim X$: P sent a message containing the statement.
- $P \Rightarrow X$: P has jurisdiction over X.
- $\#(X)$: a statement X is fresh. It has not been used before.
- $P \overset{K}{\leftrightarrow} Q$: The secret material K is only usable in the communication between *P* and *Q*.
- Encryption of X with key K is denoted in the standard way: $\{X\}_K$.
- $K_{ABi-1}$: is a secret key between *A* and *B* at the i-*1* and is used to authenticate mutually at the i.
- $Snum_B$: a sequence number of *B* which represents the time of encounters with *A*. This number will be stored in *B*'s storage.

We will list the main BAN-logic's rules that we will use to prove our protocol:

- Message meaning rule: $\frac{P \mid \equiv P \overset{K}{\leftrightarrow} Q, P \triangleleft \{X\}_K}{P \mid \equiv Q \mid \sim X}$
- Nonce Verification rule: $\frac{P \mid \equiv \#(X), P \mid \equiv (Q \mid \sim X)}{P \mid \equiv Q \mid \equiv X}$
- Jurisdiction rule: $\frac{P \mid \equiv Q \Rightarrow X, P \mid \equiv Q \mid \equiv X}{P \mid \equiv X}$
- Freshness rule: $\frac{P \mid \equiv \#(X)}{P \mid \equiv \#(X, Y)}$
- Elimination rule: $\frac{P \mid \equiv Q \mid \equiv (X, Y)}{P \mid \equiv Q \mid \equiv X}$

In our protocol, there are six main goals we need to achieve. We use symbols *A* and *B* to represent two objects in an authentication process.

- **G1:** $A \mid \equiv A \xleftrightarrow{K_{ABi-1}} B$
- **G2:** $B \mid \equiv A \mid \equiv A \xleftrightarrow{K_{ABi-1}} B$
- **G3:** $B \mid \equiv A \xleftrightarrow{K_{ABi-1}} B$
- **G4:** $A \mid \equiv B \xleftrightarrow{Snum_B} A$
- **G5:** $B \mid \equiv A \mid \equiv B \xleftrightarrow{Snum_B} A$
- **G6:** $A \mid \equiv B \mid \equiv B \xleftrightarrow{Snum_B} A$

We use BAN-logic to demonstrate the key exchange scheme between two objects that have created an interaction term and shared the secure proof or the two objects need to prove theirs through an intermediary. At each transaction, an object sends a message X to another object at the same level, *A* and *B* reciprocate their roles so that we only prove towards sending requests and verifying requests. X is encrypted by the old shared-key $K_{ABi-1}$ with $Snum_B$ which the sender wants to confirm. In particular, *A* is the sender and *B* is the receiver. The above six goals are based on the results that the key exchange protocol desires. In **G1** and **G3**, *A* and *B* both believe the shared key information *A* has been successfully established with B, since **G2** provides a strong proof that *B* knows that *A* is assured about their shared material. Similar to *Snum*, however, there is a difference in the key exchange process; the sender always transmits the receiver's *Snum* along with the encrypted message in which *Snum* represents the number of successful interactions between *A* and *B*. Thus, in **G5** and **G6**, *A/B* makes sure that *B/A* assures about the *Snum* value of the receiver, which is protected.

$$X = <A \xleftrightarrow{K_{ABi-1}} B, B \xleftrightarrow{Snum_B} A, M, ID>$$

1. Assumptions concerning the initial state are written as

- *A1:* $B| \equiv B \xleftrightarrow{K_{ABi-1}} A$
- *A2:* $B| \equiv A \Rightarrow A \xleftrightarrow{K_{ABi-1}} B$
- *A3:* $B| \equiv \#ID$
- *A4:* $A| \equiv \#M$
- *A5:* $A| \equiv B \Rightarrow B \xleftrightarrow{Snum_B} A$

2. Analysing our proposed scheme based on the BAN-logic rules and assumptions about authenticating among the mobility objects follow the above main goals.

- Because *A* and *B* have created $K_{ABi-1}$ after they establish a secure encounter, it means that *A* believes that *A* has shared the secret shared-key $K_{ABi-1}$ with *B*; therefore, we achieve **G1**.
- In accordance with *A1* and the encrypted message *X*, applying the message-meaning rule to derive:

$$\frac{B| \equiv B \xleftrightarrow{K_{ABi-1}} A, B \triangleleft \{X\}_{K_{ABi-1}}}{B| \equiv A| \sim X} \quad (4)$$

- According to *A3*, applying the freshness rule to derive:

$$\frac{B| \equiv \#(ID)}{B| \equiv \#(X)} \quad (5)$$

- In accordance with (4) and (5), we apply the nonce-verification rule to receive:

$$\frac{B| \equiv \#(X), B| \equiv A| \sim X}{B| \equiv A| \equiv X} \quad (6)$$

- In accordance with (6), we apply elimination rule to derive:

$$\frac{B| \equiv A| \equiv X}{B| \equiv A| \equiv A \xleftrightarrow{K_{ABi-1}} B}(G_2) \quad (7)$$

$$\frac{B| \equiv A| \equiv X}{B| \equiv A| \equiv A \xleftrightarrow{Snum_B} B}(G_5) \quad (8)$$

- According to *G2* and *A2*, applying jurisdiction rule to evolve:

$$\frac{B| \equiv A \Rightarrow A \xleftrightarrow{K_{ABi-1}} B, B| \equiv A| \equiv A \xleftrightarrow{K_{ABi-1}} B}{B| \equiv A \xleftrightarrow{K_{ABi-1}} B}(G_3) \quad (9)$$

- The combination of the *G1* and message-meaning rule, we can infer:

$$\frac{A| \equiv A \xleftrightarrow{K_{ABi-1}} B, A \triangleleft \{X\}_{K_{ABi-1}}}{A| \equiv B| \sim X} \tag{10}$$

- According to **A4**, applying the freshness rule to derive:

$$\frac{A| \equiv \#(M)}{A| \equiv \#(X)} \tag{11}$$

- In accordance with (10) and (11), we apply nonce-verification rule to deduce:

$$\frac{A| \equiv \#(X), A| \equiv B| \sim X}{A| \equiv B| \equiv X} \tag{12}$$

- From (12), we combine with elimination rule to derive:

$$\frac{A| \equiv B| \equiv X}{A| \equiv B| \equiv B \xleftrightarrow{Snum_{B'}} A}(G_6) \tag{13}$$

- In accordance with **G6** and **A4**, applying the jurisdiction rule to infer:

$$\frac{A| \equiv B \Rightarrow B \xleftrightarrow{Snum_B} A, A| \equiv B| \equiv B \xleftrightarrow{Snum_B} A}{A| \equiv B \xleftrightarrow{Snum_B} A}(G_4) \tag{14}$$

3. In accordance with **G1, G2, G3, G4, G5, and G6**, we can suppose that *B* believes that *A* believes they have a good shared-key $K_{ABi-1}$ in their connection. Additionally, *A* (*B*) also believes that *B* (*A*) believes $Snum_B$ is a good share-material between them.

**Threat Model and Further Security Analyses**

In our threat model, we assume that the Guide and the helpers who are already in the secured zone. Specifically, we consider that the adversary A can launch the following attacks in the newcomer (N). When the Guide (G) receives N's request, he is considered as trusted with N, but N is not. Their communications take place over insecure channels or they can employ the public channel:

- *Replay attack (Authentication)*: The adversary A will capture the previous message in the transmission channel and fraudulently has it delayed or repeated (relay them without modification). The out-of-date messages are sent to G. If G cannot check the valid and correctness of their messages, A can be involved in a secure zone.
- *Impersonation attack (Authentication):* an attack in which the adversary A tries to masquerade as a legitimate (as N) to send a valid request to G and want to be involved in the secure zone illegally.
- *Message modification attack*: the adversary A captures the messages from N and attempt to change the content of the message.
- *Man-in-the-Middle (Integrity, Confidentiality)*: when G and N have an interaction on public channels. The adversary A can eavesdrop, or delete the exchanged messages during the transmission in order to tamper the communicated data.

Then, we will analyse in detail based on the above "Threat Model" and apply the proposed protocol to prevent those attacks. Based on many known attacks, our scheme could detect, against and limit the risk of creating the lowest possible corruption so that our protocol meets the security goals mentioned in the section. We also suppose that the communication channel between the newcomer and the guide is open; anybody can eavesdrop and capture the message packages transmitted in this channel.

*Impersonation and Modification Attack:* When a device moves into an unsafe environment, unfortunately, it is attacked and masquerades some unprotected information inside its memory to make false positives or may join in unauthorized zones. There are two cases:

- When an outsider forges a relationship with a guide: they have made an interaction with each other. In our special key exchange model, they will distribute the following keys against the number of times their messages sent, which includes the sequence number and the shared key $K_{Abi-1}$. $K_{Abi-1}$ stored in encrypted even though a fake outsider acknowledges the sequence number.
- When an outsider does not know about X's Guide: It is also the goal to authenticate an object to the secure zone. At this point, we will rely on the reputation score and the number of responses to make the final decision on this object. However, Ginny will simultaneously detect proximity friends and send a list contains these friends to Newton. Newton will send the evidence of interaction with someone in the friend's list. Ginny receives and sends to the chosen helper, especially the helper that Newton sends $E_M = (K \{Newton, helpers\}||Snum = α)$ Ginny cannot read this content, will also be added to Ginny's original message to the helper. If one of $K$ and $Snum$ has problems, the helper will automatically subtract until the reputation reaches 0.

*Man-in-the-Middle:* Authentication occurs on an insecure channel between the outsider and the insider. Although the two objects are real, it may appear the eavesdropper who stands in the middle to tap or catch confidential information. Moreover, they can replace another message. The transaction between the two objects will apply to the key exchange model, the query format is $<ID | E_M>$ where $E_M$ is an encrypted message from the original one combined with the corresponding sequence number. $E_M$ is encrypted initially with an asymmetric key then it will be encrypted using a symmetric key. Although the asymmetric encryption is extremely complex and secure, it takes a lot of time to decrypt while weak objects cannot complete the exchange process. Therefore, we use both types of cryptography with sequence numbers to save time and safer.

*Replay Attack:* The intruder sends to the target the same message, which has already used in the victim's communication. The attacker either eavesdropped a message between two objects before or knew the message format from his previous communication with one of the sides. The transmission message contains shared-keys to use for mutual authentications. However, it is encrypted with the previous shared-key, which only two objects can understand.

**Security Features Comparison**

To illustrate the effectiveness of our proposed scheme, we have presented the comparison of security features provided by our protocol and two related works by evaluating whether the 12 features in Table 7. Considering the earlier criteria-related study [25, 26, 27], based on a set of 11 features from [2] and a set of 10 features which is defined in [3], we put a list of 12 following independent criteria to show that our scheme provides higher security level than the previous works. Particularly, [2, 3] did not mention the resistance of man-in-the-middle attack or not following the trace of the object as it moves from a zone to another. Moreover, we have established an authentication scheme without password ($SF_{12}$) to avoid unwanted attacks on passwords or the risk of losing passwords, though two articles mentioned password attacks. From Table 7, we can see that [2, 3], the object untraceability has not been considered. An adversary is possible to catch two communication messages on the unsecure channel and tries to expose identity of these objects. In our communication message M (Figure 2), it is encrypted in two ways that we presented in the "Key Exchange Scheme" subsection. Depending on the session key and the sequence number, M will be different and unique in each session. Therefore, our scheme also resists the object untraceability attack ($SF_{11}$). Next is the man-in-the-middle attack (MITM), the intent is the reuse of information or duplicate messages for nefarious purposes while MITM is all of the possibilities that an attacker can perform when they capture the message: observing, eavesdropping, modifying, exposing data in packets…. The ability to prevent MITM attack ($SF_5$) of our scheme is explained in the above "Threat Model" subsection. The proposed solution in [3] does not also support the features $SF_7$, $SF_{10}$ while we mentioned earlier that we only accept requests when its format is appropriate or after the voting process has been successfully accomplished. In addition, we also provide two confidential information: a session key and a sequence number to monitor and isolate each session to avoid tampering or prevent bad requests from DoS attacks ($SF_7$). In brief, our proposed scheme presented in this article can prevent several attacks and the security criteria which are aggregated and choosed appropriate features ($SF_1$-$SF_{11}$) for our article are listed in Table 7.

Table 7: Comparison of security features

| Scheme | $SF_1$ | $SF_2$ | $SF_3$ | $SF_4$ | $SF_5$ | $SF_6$ | $SF_7$ | $SF_8$ | $SF_9$ | $SF_{10}$ | $SF_{11}$ | $SF_{12}$ |
|---|---|---|---|---|---|---|---|---|---|---|---|---|
| [2] | ✓ | ✓ | ✓ | ✓ | ✗ | ✓ | ✓ | ✓ | ✓ | ✓ | ✗ | ✗ |
| [3] | ✓ | ✓ | ✓ | ✓ | ✗ | ✓ | ✗ | ✓ | ✓ | ✗ | ✗ | ✗ |
| Ours | ✓ | ✓ | ✓ | ✓ | ✓ | ✓ | ✓ | ✓ | ✓ | ✓ | ✓ | ✓ |

Note: SF: Security Features; $SF_1$: session key verification; $SF_2$: no key/password exposure; $SF_3$: provision of key exchange; $SF_4$: replay attack; $SF_5$: man-in-the-middle attack; $SF_6$: impersonation attack; $SF_7$: DoS attack; $SF_8$: object anonymity; $SF_9$: mutual authentication; $SF_{10}$: perfect forward secrecy; $SF_{11}$: object untracebility attack; $SF_{12}$: passwordless authentication.
✓ is a scheme achieves the corresponding feature while ✗ is not support or not mentioned in the article.

## Conclusions

Several IoT scenarios include aspects of mobile objects that face several encounters at different secure zones. These locations can be a meeting room, a building, an area or a city based on the network is established. For encounters with strangers, the objects meet various security challenges and the flexibility of authentication to participate in secured zones with its resource constraints. In this article, we introduced a new approach to mutual authentication that includes a reputation-based evaluation and a key exchange mechanism to lower the costs effectively between the two objects. In the second phase, we described the term of reputation based on friendships as a trustworthy value. At the same time, we proposed the key exchange mechanism based on the number of previous communications to replenish for the process, each sequence the cryptographic key would be different. In addition, the scheme selected the relevant modern cryptosystems to improve the effectiveness of information security. Our approach solved the problem of previous works that depends on a CA trust model, the execution time, the cost of consumption, energy-efficient solutions and the limit the number of passing devices. We devised a simulation based on the dataset of CRAWDAD combined with our additional dataset, which created a pretty big and messy script for more realistic testing. Our formulas are just average operations in which the dividend is not large. Thus, the execution time is quite small for the whole large script. Besides, we compared the results of two previous secure authentication protocols that our algorithm is a lower communication cost. To illustrate the effectiveness and establish the practical value of our proposed scheme, we have also presented the comparison of security features provided by our protocol and related works by employing an objective third-party evaluation metric [26, 28] The analyses have shown that our newly proposed scheme is very promising, it can prevent typical attacks and fulfill all the security features as being concerned in that evaluation metric.

To develop a large scenario like IoT, the number of objects that displace to other authorized areas quickly, the other security issues such as privacy for objects are still the challenges. Future work will create a large set of data that works on the modified protocol. In addition, the association of the authentication process and the different kinds of privacy protection [29, 30] are possibilities to support to extend the security capabilities of our protocol, especially with resource-constrained IoT devices. In [31], for example, the authors proposed a scenario related to our protocol that applies cryptographic-based protocol to satisfy desired security requirements of electronic voting and protect privacy, so we can extend the security of the proposed protocol to improve its practical value. The dynamic scenario we proposed would affect the access request since the authentication process begins getting the wanderer's joining request. A new approach [32] will be greater flexibility and availability for our system and still ensures security for circumstances can happen in the wireless IoT network system. Furthermore, our mechanism had not really paid enough attention to the real-time issue of the authentication and the speed of mobility objects. To continue improving and expanding the scope, we will also consider the real-time data access scenario, or the intervals of friendships to ensure that it remains reliable [26, 33, 34].

## Data Availability

The datasets generated during and/or analysed during the current study are available in the Zenodo repository, **https://doi.org/10.5281/zenodo.1583282**.

## Acknowledgment

This research is funded by Vietnam National University Ho Chi Minh City (VNU-HCM) under grant number B2018-20-08. We also thank DSTAR Lab members for their meaningful help during this manuscript preparation. We also appreciate all great comments from the reviewers as well as the editor in charge to help us improve this manuscript.